\documentclass[prd,twocolumn]{revtex4}
\usepackage{graphicx, epsfig}
\usepackage{color}
\usepackage{mathrsfs}
\usepackage{amsmath}

\newcommand{\be}{\begin{equation}}
\newcommand{\ee}{\end{equation}}
\newcommand{\bea}{\begin{eqnarray}}
\newcommand{\eea}{\end{eqnarray}}

\newcommand{\gapp}{\mathrel{\raise.3ex\hbox{$>$}\mkern-14mu
              \lower0.6ex\hbox{$\sim$}}}
\newcommand{\lapp}{\mathrel{\raise.3ex\hbox{$<$}\mkern-14mu
              \lower0.6ex\hbox{$\sim$}}}

\begin{document}
\pdfoutput=1

\title{Time Evolution of Temperature and Entropy of a Gravitationally Collapsing Cylinder}
\author{Evan Halstead}
\author{Peng Hao}
\affiliation{HEPCOS, Department of Physics, SUNY at Buffalo, Buffalo, NY 14260-1500} 

\begin{abstract}
We investigate the time evolution of the temperature and entropy of a gravitationally collapsing cylinder, represented by an infinitely thin domain wall, as seen by an asymptotic observer.  Previous work has shown that the entropy of a spherically symmetric collapsing domain approaches a constant, and we follow this procedure using a (3+1) BTZ metric to see if a different topology will yield different results.  We do this by coupling a scalar field to the background of the domain wall and analyzing the spectrum of radiation as a function of time.  We find that the spectrum is quasi-thermal, with the degree of thermality increasing as the domain wall approaches the horizon.  The thermal distribution allows for the determination of the temperature as a function of time, and we find that the late time temperature is very close to the Hawking temperature and that it also exhibits the proper scaling with the mass.  From the temperature we find the entropy.  Since the collapsing domain wall is what forms a black hole, we can compare the results to those of the standard entropy-area relation.  We find that the entropy does in fact approach a constant that is close to the Hawking entropy.  However, the time dependence of the entropy shows that the entropy decreases with time, indicating that a (3+1) BTZ domain wall will not collapse spontaneously.  
\end{abstract}
\maketitle

\section{Introduction}
It is well known, based primarily on the work of Bekenstein, Gibbons, and Hawking, that the entropy of a black hole is proportional to its area and that even though supposedly nothing can escape from within a black hole, quantum fluctuations near the event horizon will produce a spectrum of radiation that is thermal \cite{Hawking:1974sw, Hartle:1976tp, Gibbons:1976ue}.  Since then, much work has been done to reproduce this result with theories of quantum gravity, but most of these theories do not analyze the time evolution of the system.  The conventional process to determine the entropy uses the Bogolyubov method to first determine the temperature and from there find the entropy.  This, however, only utilizes the initial and final states of the system, and therefore there is no knowledge of the time dependence.  Recently, Vachaspati and Stojkovic developed a quantum treatment to determine the quantum radiation given off during gravitational collapse in a time dependent manner, then Greenwood expanded on this to determine the time dependence of the entropy \cite{Vachaspati:2006ki, Vachaspati:2007hr, Greenwood:2008vu, Greenwood:2008zg, Wang:2009ay, Greenwood:2010mr}.  These papers used a spherically symmetric collapsing domain wall for their analysis, and we wish to augment their work by seeing if a different topology will yield significantly different results.  Specifically, we wish to determine the time dependence of the entropy of a gravitationally collapsing cylinder, which we will represent with an infinitely thin domain wall.  Ref.\cite{Greenwood:2009gp} used a (3+1) BTZ metric to determine the equation of motion of the collapsing cylindrical domain wall, and we will use that to complete the analysis of its thermodynamic properties.  To do this, we will first determine the wavefunctional of a scalar field coupled to the background of the collapsing shell.  Then, using the t=0 wavefunctional as a basis, we will determine the occupation number as a function of frequency.  This in turn will allow for the determination of the thermodynamic quantity $\beta$ and therefore the temperature.  We then determine the late time temperature as a function of horizon radius.  Using the thermodynamic definition of entropy, $dS=dQ/T$, we integrate to find entropy as a function of temperature, and since we know the temperature as a function of time we can find the entropy as a function of time.  We compare our results against the standard Hawking results, and we comment on the results.  

\section{Radiation Wavefunctional}

First, we will consider the radiation given off by the  cylindrical domain wall during gravitational collapse.  This will be done by coupling a scalar field, which we will decompose as
\begin{equation}
\Phi = \sum_{k} a_k(t)u_k(r),
\end{equation}
to the background of the collapsing shell.  The exact form of $u_k(r)$ will not be important to us.  To find the modes $a_k(t)$, we will insert metrics (see Ref.\cite{Greenwood:2009gp})
\begin{equation}
ds^2 = \frac{\Lambda}{3}r^2dT^2+\frac{1}{-\frac{\Lambda}{3}r^2}dr^2+r^2(d\phi^2+dz^2), r<R(t)
\end{equation}
and 
\begin{eqnarray}
ds^2 &=& -\left (-\frac{\Lambda}{3}r^2-\frac{4GM}{r}\right)dt^2+\frac{1}{-\frac{\Lambda}{3}r^2-\frac{4GM}{r}}dr^2 \nonumber \\
&&+r^2(d\phi^2+dz^2), r>R(t)
\end{eqnarray}
 into the action 
\begin{equation}
S_\Phi = \int d^4x \sqrt{-g} g^{\mu\nu} \partial_\mu \Phi \partial_\nu \Phi.
\end{equation}  
We will define
\begin{eqnarray}
f=-\frac{\Lambda}{3}R^2-\frac{4GM}{R} \\
A=-\frac{\Lambda}{3}R^2.
\end{eqnarray}
The total action will be written as the sum
\begin{equation}
S = S_{in} + S_{out},
\end{equation}
where 
\begin{equation}
S_{in} = \pi \int dt dz \int_0^{R(t)}dr r^2 \left (-\frac{(\partial_t \Phi)^2}{\dot{T}A} + \dot{T}(\partial_r \Phi)^2\right )
\end{equation}
is the metric inside the shell, 
\begin{equation}
S_{out} = \pi \int dt dz \int_{R(t)}^\infty dr r^2 \left (-\frac{(\partial_t\Phi)^2}{f}+f(\partial_r\Phi)^2\right )
\end{equation}
is the metric outside the shell, and
\begin{equation}
\frac{dT}{dt} = \frac{1}{A} \sqrt {Af-\frac{\left [A-f\right] \dot{R}^2}{f}}.
\end{equation}
According to Ref.\cite{Greenwood:2009gp}, 
\begin{equation}
\dot{R} = f \sqrt{1-\frac{fR^4}{h^2}}
\end{equation}
where 
\begin{equation}
h = \frac{f^{3/2}R^2}{\sqrt{f^2-\dot{R}^2}},
\end{equation}
so dT/dt can be rewritten as 
\begin{equation}
\frac{dT}{dt} = \frac{f}{A}\sqrt{1+\left [ A - f \right ] \frac{R^4}{h}}.
\end{equation}

In the region of interest, where $R\rightarrow R_H$ and therefore $f\rightarrow 0$, the kinetic term in $S_{in}$ dominates over that in $S_{out}$ while the gradient term in $S_{in}$ is subdominant to that in $S_{out}$.  This yields the approximate action
\begin{equation}
S \approx \pi \int dt \int dz \left [ -\int_0^{R_H} dr r^2 \frac{(\partial_t \Phi)^2}{f} + \int_{R_H}^{\infty} dr r^2 f (\partial_r \Phi)^2 \right ].  
\end{equation}
Substituting the expansion for the scalar field $\Phi$ produces 
\begin{eqnarray} \label{eq:scalaraction}
&S& = \int dt \left[ -\pi \int dz \int_0^{R_H} dr r^2 \frac{\sum_{k,k'} \dot a_k(t)\dot a_{k'}(t)u_k(r)u_{k'}(r)}{f} \right . \nonumber \\
&& \left . + \pi \int dz \int_{R_H}^\infty dr r^2 f \sum_{k,k'} a_k(t)a_{k'}(t)u'_k(r)u'_{k'}(r)\right ].
\end{eqnarray}
  Now let
\begin{equation}
M_{kk'} = 2\pi \int dz \int_0^{R_H} dr r^2 u_k(r) u_{k'}(r)
\end{equation}  
  and
\begin{equation}
N_{kk'} = - 2\pi \int dz \int_{R_H}^\infty dr r^2 f(r) u'_k(r) u'_{k'}(r),
\end{equation}   
 which allows us to rewrite $(\ref{eq:scalaraction})$ as
 \begin{equation} \label{eq:scalaraction1}
 S = \sum_{k,k'} \int dt \left [ -\frac{1}{2f}\dot a_k M_{kk'} \dot a_{k'} - \frac{1}{2} a_k N_{kk'} a_{k'}\right ].
 \end{equation}
   The action $(\ref{eq:scalaraction1})$ corresponds to the Hamiltonian
   \begin{equation}
   H = \frac{1}{2}f\Pi_k M_{kk'}^{-1} \Pi_{k'} + \frac{1}{2} a_k N_{kk'} a_{k'},
   \end{equation}
 where
 \begin{equation}
 \Pi = \frac{\partial L}{\partial \dot a}.
 \end{equation}
   
We can now find the wave function $\psi$ from
\begin{equation}
H\psi = i \frac{\partial \psi}{\partial t}.
\end{equation}
Since $\textbf M$ and $\textbf N$ are Hermitian matrices, it is possible to do a principle axis transformation to diagonalize them simultaneously.  The exact form of them, however, is not important here.  The Schr$\mathrm{\ddot {o}}$dinger Equation for a single mode will then be
\begin{equation} \label{eq:hamiltonian}
\left [-\frac{1}{2m}f\frac{\partial^2}{\partial b^2}+\frac{1}{2} K b^2 \right ] \psi (b,t) = i \frac{\partial \psi (b,t)}{\partial t}, 
\end{equation}
where $\textit m$ and $\textit K$ are the eigenvalues of $\textbf M$ and $\textbf N$, respectively, and $\textit b$ is the eigenmode.

This, however, only represents the Hamiltonian of the radiation, not the total system.  The Hamiltonian of the system can be written as
\begin{equation}
H_{total}=H_{wall}+H_{rad}.
\end{equation}
Ref.\cite{Greenwood:2009gp} showed that 
\begin{equation}
H_{wall}^2 = (f \Pi_R)^2 + f(2\pi \mu R^2)^2,
\end{equation}
where $\mu=\sqrt{\Lambda R_H^2/3}-2\pi \sigma G R_H$ and $\sigma$ is the tension in the wall.  As $R\rightarrow R_H$, $f\rightarrow 0$, and the first term on the right hand side will dominate (since $\Pi \sim f^{-3/2}$), yielding
\begin{equation}
H_{wall} \approx -f\Pi_R,
\end{equation}
where we have chosen the negative solution because the shell is collapsing.  
The Hamiltonian of the radiation is
\begin{equation}
H_{rad} = \frac{f}{2m}\Pi_b^2+\frac{K}{2}b^2.
\end{equation}
The Hamiltonian of the entire system is therefore
\begin{equation} \label{eq:Htotal}
H_{total}=-f\Pi_R+\frac{f}{2m}\Pi_b^2+\frac{K}{2}b^2.
\end{equation}
Note that the momentum operators have different indices, where $\Pi_R=-i\partial / \partial R$ is the momentum operator conjugate to the shell's position and $\Pi_b=-i\partial / \partial b$ is the momentum operator conjugate to the eigeinmode $\textit b$.  The wavefunction of the entire system is therefore a function of $\textit b$, $\textit R$, and $\textit t$, which we will write as
\begin{equation}
\Psi = \Psi (b,R,t).
\end{equation}
Using ($\ref{eq:Htotal}$), the Functional Schr$\mathrm{\ddot{o}}$dinger equation can be written as
\begin{equation}\label{eq:functionalschr}
if\frac{\partial \Psi}{\partial R} - \frac{f}{2m}\frac{\partial^2\Psi}{\partial b^2}+\frac{K}{2}b^2\Psi = i\frac{\partial \Psi}{\partial t}.
\end{equation}
Solving this requires knowing the classical equation of motion for the distance of the domain wall, which we will take as approximately
\begin{equation}\label{eq:Rdot}
\dot R \approx -f,
\end{equation}
where again the negative sign was chosen because the shell is collapsing.  Since R(t) is only a function of time, we can rewrite ($\ref{eq:functionalschr}$) as 
\begin{equation}
if\frac{1}{\dot R}\frac{\partial \Psi}{\partial t}-\frac{f}{2m}\frac{\partial^2\Psi}{\partial b^2}+\frac{K}{2}b^2\Psi=i\frac{\partial \Psi}{\partial t}.
\end{equation} 
According to ($\ref{eq:Rdot}$), however, $f/\dot{R}=-1$, therefore
\begin{equation}
-\frac{f}{2m}\frac{\partial^2\Psi}{\partial b^2}+\frac{K}{2}b^2\Psi = 2i\frac{\partial \Psi}{\partial t}.
\end{equation}
Rewriting this in the standard simple harmonic oscillator form yields
\begin{equation}
\left[-\frac{1}{2m}\frac{\partial^2}{\partial b^2}+\frac{m}{2}\omega^2 b^2\right ] \Psi(b,\eta) = i\frac{\partial \Psi (b, \eta)}{\partial \eta},
 \end{equation}
 where 
 \begin{equation}
 \omega^2=\frac{K}{mf}=\frac{\omega_0^2}{f} 
 \end{equation}
 and
 \begin{equation}
 \eta=\frac{1}{2}\int_0^t dt' f.
 \end{equation}
 Here we have chosen $\eta(t=0)=0$.  

To proceed further, we need to incorporate the equation of motion R(t) given by (see Ref.\cite{Greenwood:2009gp})
\begin{equation}
R(t)=R_h+(R_0-R_h)e^{\Lambda R_h t/3},
\end{equation}
where we are assuming a negative cosmological constant.  
At early times, R(t) is approximately constant, so the initial vacuum state for the wavefunction $\psi$ is that of a simple harmonic oscillator given by
\begin{equation}
\psi (b, \eta =0) = \left (\frac{m \omega_0}{\pi}\right )^{1/4} e^{-m\omega_0 b^2/2}.  
\end{equation}
At later times, 
\begin{equation}
f\approx-\frac{\Lambda R_h^2}{3}-\frac{4 G M}{R(t)}, 
\end{equation}
and the exact solution for the wavefunction is (see Ref.\cite{Dantas:1990rk})
\begin{equation} \label{eq:latewavefunction}
\psi (b, \eta) = e^{i\alpha (\eta)}\left (\frac{m}{\pi \rho^2}\right)^{1/4} \exp{\left [i\frac{m}{2}\left(\frac{\rho_\eta}{\rho}+\frac{i}{\rho^2}\right)b^2\right ]},
 \end{equation}
where $\rho_\eta$ denotes the derivative of $\rho(\eta)$ with respect to $\eta$, and $\rho(\eta)$ is the solution to the equation
\begin{equation}
\frac{\partial^2 \rho}{\partial \eta^2}+\omega^2(\eta)=\frac{1}{\rho^3}
\end{equation}
 with initial conditions
 \begin{eqnarray}
 \rho(0)=\frac{1}{\sqrt{\omega_0}} \\
 \rho_{\eta}(0)=\frac{\partial \rho}{\partial \eta} \mid_0 = 0.
 \end{eqnarray}
The phase $\alpha$ is given by
\begin{equation}
\alpha(\eta)=-\frac{1}{2}\int_0^\eta \frac{d\eta'}{\rho^2(\eta')}.
\end{equation}   

\section{Occupation Number of the Radiation}

\begin{figure}[ht]
\includegraphics{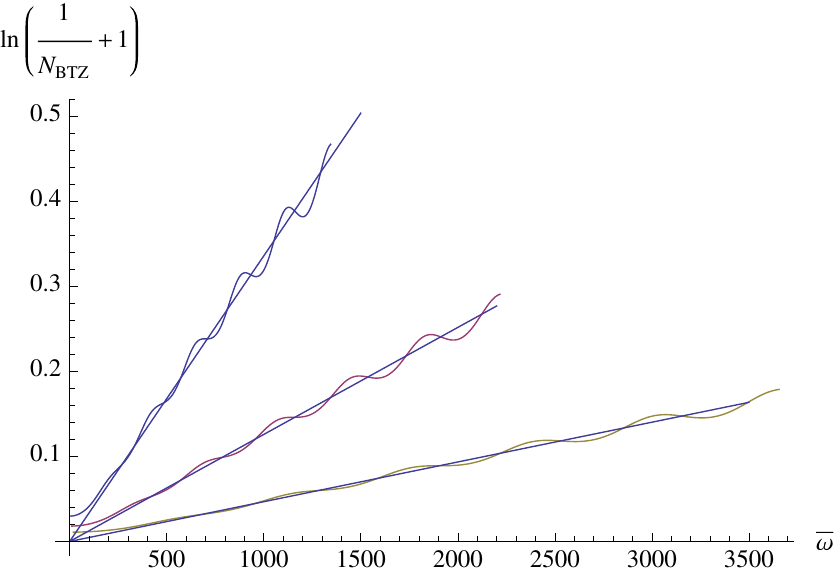}
\caption{The straight lines are the best fit to the curve and the slopes represent $\beta$.}
\label{fig:lnNbtz}
\end{figure}

\begin{figure}[ht]
\includegraphics{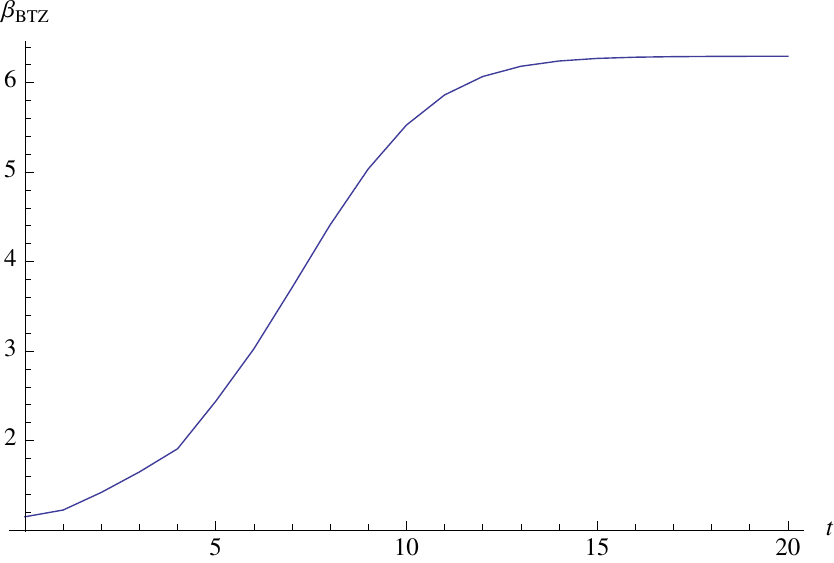}
\caption{$\beta$ as a function of time for a collapsing 3+1 BTZ domain wall.}
\label{fig:betabtz}
\end{figure}

\begin{figure}[ht]
\includegraphics{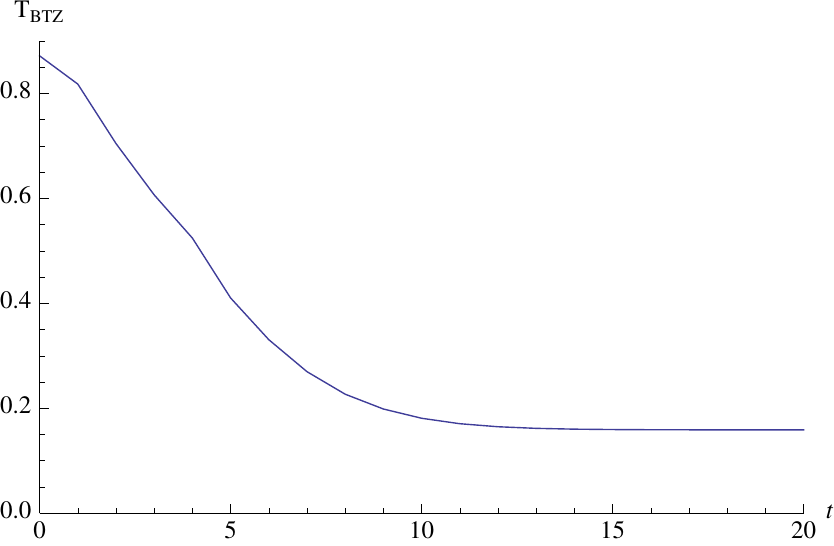}
\caption{Temperature as a function of time for a collapsing 3+1 BTZ domain wall.}
\label{fig:tempbtz}
\end{figure}

Consider an observer with detectors that are designed to register particles for the scalar field $\phi$ at early times.  At late times, the observer will interpret each mode $b$ in terms of the simple harmonic oscillator states, with final frequency $\bar \omega$.  The number of quanta in eigenmode $b$ can be found by decomposing the wavefunction $(\ref{eq:latewavefunction})$ into the simple harmonic oscillator states and evaluating the occupation number.  The wavefunction $\psi$ written in terms of the simple harmonic basis, $\{\varphi_n \}$, at $t=0$ is given by
\begin{equation}
\psi (b,t) = \sum_n c_n(t) \varphi_n(b),
\end{equation}
where 
\begin{equation}
c_n(t) = \int db \varphi_n^\ast(b) \psi (b,t),
\end{equation} 
which is the overlap of a Gaussian with the simple harmonic basis functions.  The occupation number at eigenfrequency $\bar \omega$ is given by the expectation value
\begin{equation}
N(t,\bar \omega) = \sum_n n |c_n|^2.  
\end{equation}
After substitution, we find that the occupation number in the eigenmode $b$ is given by (see Appendix B in Ref.\cite{Greenwood:2009pd})
\begin{equation}
N(t,\bar{\omega}) = \frac{\bar{\omega} \rho^2}{4}\left [ \left (1-\frac{1}{\bar{\omega} \rho^2}\right )^2 + \left (\frac{\rho_t}{f\bar{\omega} \rho}\right )^2\right ].
\end{equation}
\begin{figure}[ht]
\includegraphics{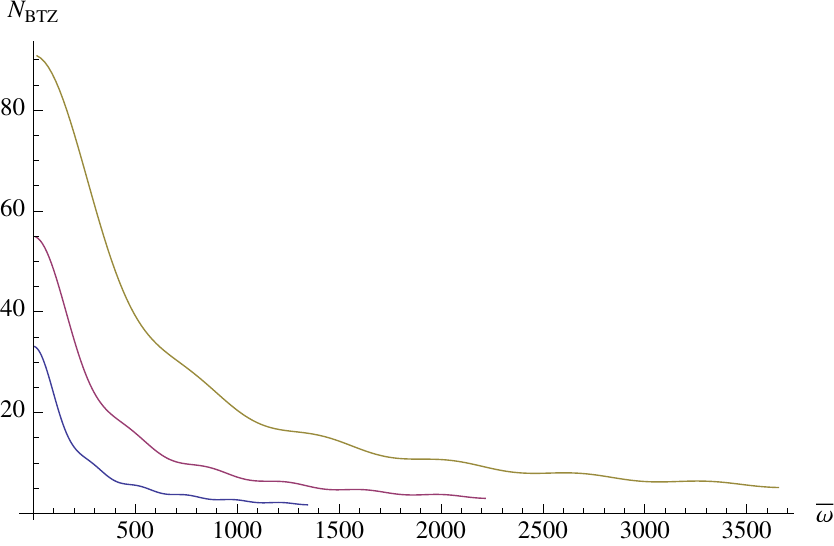}
\caption{Occupation number of the radiation as a function of frequency.  From bottom to top, the three lines are for t=13, t=14, and t=15, respectively.}
\label{fig:Nbtz}
\end{figure}
The plot in Fig.(\ref{fig:Nbtz}) shows the occupation number as a function of frequency for three time slices.  
This is very similar to a typical Planck distribution.  Therefore, we can find the temperature of the radiation by comparing the occupation number of each eigenmode $b$ to that of the Plank distribution,
\begin{equation}
N_P=\frac{1}{e^{\beta \bar{\omega}}-1}
\end{equation}
If one plots $\ln{(1+1/N)}$ as a function of $\bar{\omega}$, the slope will yield $\beta$, as shown in Fig.(\ref{fig:lnNbtz}).  It should be noted that this slope is conjugate to the coordinate $\eta$.  If we wish to obtain $\beta$ as a function of observer time $t$, then we must rescale according to
\be
\beta^{(t)}=\frac{2\beta^{(\eta)}}{f}.
\ee
The plot is shown in Fig.(\ref{fig:betabtz}).
By inverting $\beta$, we find the temperature as a function of time, as shown in Fig.(\ref{fig:tempbtz}).

We will now compare the late time temperature to the Hawking temperature.  According to \cite{Lemos:1994xp},
\be
T_H=\sqrt{-\frac{\Lambda}{3}}\frac{3}{4\pi}\left(\frac{M}{2}\right)^{1/3}.
\ee

\begin{figure}[ht]
\includegraphics{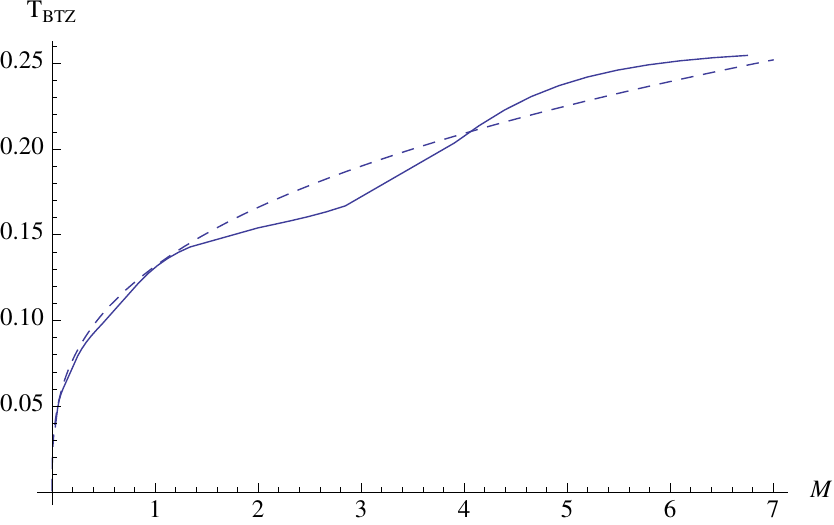}
\caption{Temperature as a function of mass for a BTZ domain wall.}
\label{fig:tempdepbtz}
\end{figure}

The ratio of the late time temperature in the plot to the Hawking temperature for $R_H=1$ is $T_{BTZ}/T_H=1.33$.  The plot of the domain wall's late time temperature versus mass, as well as the best-fit curve, is shown in Fig.(\ref{fig:tempdepbtz}).  
The equation of the best-fit curve is
\be
T_{BTZ}=0.131724 M^{1/3}
\label{eq:tempdepbtz}
\ee
We can therefore see that the late time temperature exhibits the proper scaling with the mass of the black hole, that is $T\sim M^{1/3}$.

\section{Entropy}
 The thermodynamic definition of entropy in terms of temperature is
 \begin{equation}
 S=\int{\frac{dQ}{T}}.
 \end{equation}
 Since changing the energy of the domain wall is the same as changing the mass, this can also be written as 
 \begin{equation}
 S=\int{\frac{dM}{T}}
 \label{eq:entropydefinition}
 \end{equation}
 Therefore, if we can determine how the temperature of the domain wall depends on the mass, we will be able to find an expression for the entropy in terms of the temperature.  And since we know the time evolution of the temperature, we can determine the time evolution of the entropy as well. 
 
We insert Eq.(\ref{eq:tempdepbtz}) into Eq.(\ref{eq:entropydefinition}), and after integrating we find that the entropy as a function of temperature is 
\be
S_{BTZ}=\frac{3 T^2}{2 \gamma},
\ee
where $\gamma=0.131724$.

\begin{figure}[ht]
\includegraphics{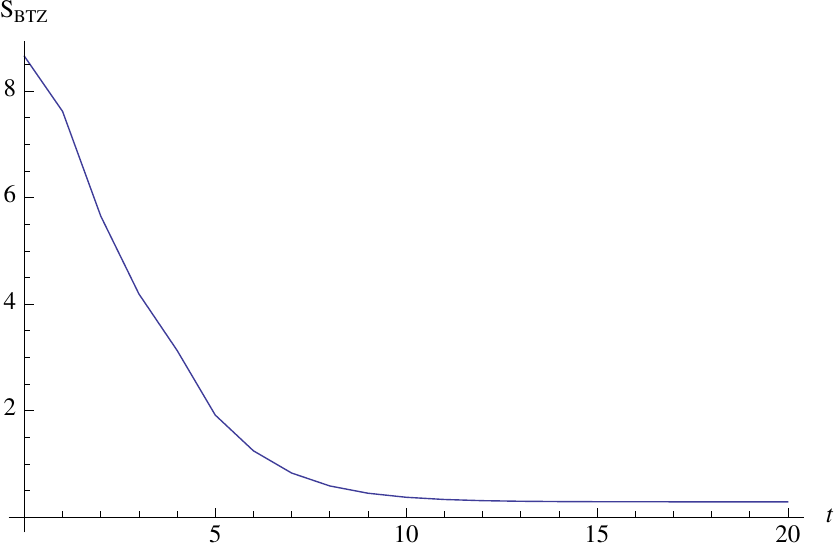}
\caption{Entropy as a function of time for a BTZ domain wall.}
\label{fig:entropybtz}
\end{figure}

The corresponding plot of entropy as a function of time is shown in Fig.(\ref{fig:entropybtz}).  
Notice that since the entropy is proportional to the square of the temperature, the entropy decreases along with the temperature.  At late times, the domain wall reaches a static state and the entropy ceases to decrease.  According to Ref\cite{Akbar:2011qw}, a BTZ black string does not evaporate once formed, and is thus stable.  However, our result of decreasing entropy seems to imply that a (3+1) BTZ domain wall would not collapse spontaneously under these conditions, so the black string would never be formed in this way.  In fact, the assumption of a collapsing cylinder may be wrong entirely, meaning that perhaps it is the expanding solution which leads to an increase in entropy.

 \section{conclusion}
We investigated the time evolution of the temperature and entropy of a 3+1 BTZ domain wall (representing a cylinder).  This was done by coupling a scalar field to the background of the domain wall and evaluating the occupation number.  From the occupation number we were able to determine $\beta$ and therefore the temperature.  We found that the temperature of the 3+1 BTZ domain wall decreased, then approached a constant, and that the late time temperature exhibited very good agreement with the Hawking temperature of a static black hole.  It also produced scaling with mass that matched the literature.  The entropy, however, exhibited very interesting behavior, as the entropy actually decreased over time.  At this point we are unable to determine the exact cause of the decrease, however, because this particular metric contains both an anti-de Sitter cosmological constant and a different topology from previously studied cases.  The interesting implication here, though, is that the collapse of a (3+1) BTZ domain wall does not happen spontaneously.  Future work could examine whether the expanding solution produces an increasing entropy.   

\begin{acknowledgements}
The authors would like to thank D. Stojkovic and E. Greenwood for their insight and their helpful conversations.
\end{acknowledgements}

\end{document}